\documentclass[11pt]{article} 
\usepackage{amsfonts,amscd,amsmath}
\def\NP#1#2{ Nucl.Phys. B#1 (#2)} 
\def\PL#1#2{ Phys.Lett. B#1 (#2)}
 
\def\MPL#1#2{ Mod.Phys.Lett.A#1 (#2)} 
 
\def\PR#1#2{Phys.Rev. D#1 (#2)} 

\def\ATMP#1#2{ Adv.Theor.Math.Phys. #1 (#2)}
\def\HP#1#2{ JHEP #1 (#2)} 
\newcommand{\adS}{\text{AdS}_{d+1}}
\newcommand{\ads}{\text{AdS}_3}
\newcommand{\bga}{\bar\gamma}
\newcommand{\bbe}{\bar\beta}
\newcommand{\ep}{\text e}
\newcommand{\pd}{\partial} 
\newcommand{\bz}{\bar z} 
\newcommand{\zo}{z_{12}} 
\newcommand{\oh}{\frac{1}{2}}
\newcommand{\ap}{a}
\title{Probing $\ads$/CFT correspondence via world-sheet methods and 2d 
gravity like scaling arguments}
\author{Oleg Andreev\thanks{Alexander von Humboldt Fellow} 
\thanks{e-mail:  andreev@physik.hu-berlin.de}
\thanks{Permanent address: Landau Institute, Moscow, Russia}
\\ \\
  Humboldt--Universit\"at zu Berlin, Institut f\"ur Physik\\
Invalidenstra\ss e 110, D-10115 Berlin, Germany}
\date{}
\begin{document}
\maketitle
\begin{abstract}
 We show how some features of the AdS/CFT correspondence for $\ads$ can easily be 
understood via standard world-sheet methods and 2d gravity like scaling arguments. 
To do this, we propose a stringy way for perturbing two-dimensional CFT's around 
their critical points. Our strategy is to start from a stringy (world-sheet) 
representation of 2d CFT in space-time. Next we perturb a world-sheet action by 
some marginal operators such that the space-time symmetry becomes finite 
dimensional. As a result, we get a massive FT in space-time with a scale provided by 
two-dimensional coupling constant. It turns out that there exists a perturbation 
that leads to string theory on $\ads$. In this case the scale is equivalently provided by the 
radial anti-de-Sitter coordinate.
\\
  PACS  number(s): 11.25.Hf, 11.25.Pm, 11.15.Pg
\end{abstract}

\vspace{-12cm}
\begin{flushright}
hep-th/9909222\\
HU Berlin-EP-99/55
\end{flushright}

\vspace{11cm}

%_______________________     I N T R O D U C T I O N    _________________

\section{Introduction and Review} 
\renewcommand{\theequation}{1.\arabic{equation}}
\setcounter{equation}{0}
Relations between gauge fields and strings present an old, fascinating and 
unanswered question (see, e.g. \cite{polbook}). In the last two years there has been 
much progress in understanding some sort of holographic correspondence, the 
so-called AdS/CFT correspondence, between 
superconformal Yang-Mills theory and supergravity or string theory on 
anti-de-Sitter spaces $\adS$ (for a review and refs., see,  \cite{adsrev}). 
Equivalence between theories in different dimensions raises questions about how 
detailed bulk information in one theory can be completely coded in lower dimensional 
degrees of freedom. Despite the large amount of evidence for the 
AdS/CFT correspondence, there is not yet any direct translation of the configuration 
of one theory to the other. At the present time it is not known whether the situation 
may be taken under control. The purpose of this paper is to provide more evidence in 
favour of such the correspondence for string propagation on curved space-time 
manifolds that include $\ads$.

There is good motivation for specializing to $\ads$ \footnote{It should be noted 
that the issue of the string propagation on $\ads$ is an old story (for a recent review of 
this issue see \cite{mario} and references therein).}. First, in this case CFT is two-dimensional, 
so the corresponding conformal symmetry is 
infinite dimensional.  In general, two dimensional CFT's and perturbations around 
them are better understood than their higher dimensional analogues and one may hope 
that this will also be the case here. Second, string theory on $\ads$ can be defined 
without turning on RR fields and thus should be more amenable to standard 
world-sheet methods. So we are bound to learn something if we succeed.

The features of the AdS/CFT correspondence that are most relevant to our discussion 
are the following:

{\it  1.1 AdS geometry}. We use the Euclidean version of $\ads$. In this case the metric
 is given by 
\begin{equation}\label{mr}
ds^2=\frac{l^2}{{\mathbf r}^2}(d{\mathbf r}^2
+d{\boldsymbol\gamma }d{\boldsymbol\bga} )\quad,
\end{equation}
where $l$ is the radius of $\ads$. Another convenient set of coordinates is 
$({\boldsymbol\varphi}, {\boldsymbol\gamma}, {\boldsymbol\bga})$ with 
${\mathbf r}=\ep^{-{\boldsymbol\varphi}}$. In these coordinates the metric is
\begin{equation}\label{mf}
ds^2=l^2(d{\boldsymbol\varphi}^2+
\ep^{2{\boldsymbol\varphi}}d{\boldsymbol\gamma} d{\boldsymbol\bga} )\quad.
\end{equation}
The boundary consists of a copy of ${\mathbf R}^2$ at ${\mathbf r}\rightarrow 0$, 
which in the 
$({\boldsymbol\varphi} ,{\boldsymbol\gamma} ,{\boldsymbol\bga})$ coordinates 
corresponds to ${\boldsymbol\varphi}\rightarrow +\infty$, 
together with a single point at ${\mathbf r}\rightarrow+\infty$, or 
${\boldsymbol\varphi} \rightarrow -\infty$. 
Thus, in this representation, the boundary of $\ads$ is obtained by adding a point at 
infinity to ${\mathbf R}^2$, which is nothing but a sphere ${\mathbf S}^2$.

{\it  1.2 World-sheet description.} Let ${\mathbf S}^2$ be a world-sheet whose 
coordinates are 
$(z,\bz)$ and $(\varphi , \gamma ,\bga )$ be sigma model 
quantum fields on it \footnote{We use bold letters for the space-time notation
here and below.}. After introducing a NS $B_{\mu\nu}$ field that is 
necessary for conformal invariance the sigma model world-sheet Lagrangian 
with the metric $G_{\mu\nu}$ defined by \eqref{mf} is given by
\begin{equation}\label{sig1}
{\cal L}\sim \frac{l^2}{l_s^2}\bigl(\pd\varphi\bar\pd\varphi+
\ep^{2\varphi}\pd\bga\bar\pd\gamma\bigr )(z,\bz )\quad.
\end{equation}
Here $l_s$ is the fundamental string length. In fact, the above construction defines 
the embedding: ${\mathbf S}^2\rightarrow\ads$. From the sigma model point of view 
one can consider the zero modes of theses fields as the coordinates in space-time, 
i.e. the zero modes of $(\varphi , \gamma ,\bga )$ are 
$({\boldsymbol\varphi}, {\boldsymbol\gamma}, {\boldsymbol\bga})$. The latter has a 
further consequence that we will exploit in section 2.

It is convenient to introduce a pair of auxiliary fields $(\beta,\bar\beta)$, and rewrite 
the Lagrangian as 
\begin{equation}\label{sig2}
{\cal L}\sim \frac{l^2}{l_s^2}\bigl(\pd\varphi\bar\pd\varphi+\beta\bar\pd\gamma+
\bar\beta\pd\bga -\beta\bar\beta\ep^{-2\varphi}\bigr)(z,\bz )\quad.
\end{equation}
Taking into account a proper measure renormalization as well as rescaling of the fields, 
at the quantum level one gets the SL(2) WZW action with 
${\text k}=l^2/l^2_s$ \cite{adsrev, mario}. 

At this point, it is necessary to make a couple of remarks. First, at the quantum level 
the last term in \eqref{sig2} becomes the screening operator of the SL(2) WZW 
model \footnote{Strictly speaking, it is the product of the holomorphic and 
anti-holomorphic screening operators.}. Second, at the classical level this term drops in 
the limit $\varphi\rightarrow+\infty$ that results in the free actions. 
On the other hand, we have seen that this limit is nothing but the definition of the 
boundary component $\mathbf R^2$ for $\ads$. So at the classical level 
we have for the space-time 
theory the following picture: string theory in the background whose geometry coincides 
with the boundary of $\ads$ is described in terms of the world-sheet free fields while 
string theory in the background whose geometry coincides with the bulk of $\ads$ can 
be thought of as a 
perturbation around this free theory. 
Is this the case for the first quantized theory? To answer this question, it is desirable 
to have new ways of understanding the description of {\it boundary in space-time} in terms of 
the world-sheet path integral whose {\it world-sheet is a closed surface}. 
In the present paper, we will suggest a way to do so. 

 {\it  1.3 AdS/CFT correspondence}. Maldacena conjectured \cite{malda} (see 
\cite{adsrev} for a review and refs.) that string theory on AdS times a compact space 
is dual to a CFT at the boundary of AdS. Following this work, methods for calculating 
correlation functions 
of various operators in CFT's were proposed by Gubser, Klebanov, Polyakov, and 
Witten \cite{gkpw}. The basic idea is to identify the generating functional of 
connected Green functions in the CFT with the minimum of the classical 
string (supergravity) theory action, subject to some boundary conditions. To be more 
precise, let ${\boldsymbol\phi} (\Vec{\boldsymbol\gamma},{\mathbf r})$ be a scalar 
field in $\ads$ obeying the Laplace equation 
$(\hat\Delta+{\mathbf m}^2)\,{\boldsymbol\phi}=0$.  Let 
${\boldsymbol\phi} _0$  be the restriction of $\boldsymbol\phi $ to the boundary of 
$\ads$. The AdS/CFT correspondence assumes that the ${\boldsymbol\phi} _0$ 
should be considered to couple to a conformal operator $\boldsymbol{\cal O}$ via a 
coupling $\int {\boldsymbol\phi} _0{\boldsymbol{\cal O}} \, d^2{\boldsymbol\gamma}$. 
The ansatz is given by
\begin{equation}\label{corr}
\langle \exp\int {\boldsymbol\phi} _0{\boldsymbol{\cal O}} \rangle_{\text{CFT}}=
\exp \bigl(-{\mathbf I}_{\text{\bf s}}({\boldsymbol\phi} )\bigr)\quad.
\end{equation}
Here ${\mathbf I}_{\text{\bf s}} $ is the classical supergravity action. 

It should be noted that the asymptotic behaviour of the classical solution is given by 
\begin{equation}\label{ass}
{\boldsymbol\phi} (\Vec{\boldsymbol\gamma},{\mathbf r})\rightarrow 
{\mathbf r}^{2-\Delta}\bigl({\boldsymbol\phi} _0(\Vec{\boldsymbol\gamma})+
O({\mathbf r}^2)\bigr)+
{\mathbf r}^\Delta\bigl({\mathbf A}(\Vec{\boldsymbol\gamma})+
O({\mathbf r}^2)\bigr)\quad,
\end{equation}
where $\Delta$ is one of the roots of 
\begin{equation}\label{kpz}
\Delta (\Delta -2)={\mathbf m}^2 \quad.
\end{equation}
${\boldsymbol \phi} _0(\Vec{\boldsymbol\gamma})$ is regarded as a ``source'' 
function while ${\mathbf A}(\Vec{\boldsymbol\gamma})$ describes physical 
fluctuations. 

It is easy to see that there are two roots of \eqref{kpz} for $-1<{\mathbf m}^2<0$. 
Explicitly, 
\begin{equation}\label{roots}
\Delta_{\pm}=1\pm\sqrt{1+{\mathbf m}^2}\quad.
\end{equation}
It was understood by Breitenlohner and Freedman in the early eighties that for this
mass range there exist two possible quantizations for the scalar field while for 
${\mathbf m}^2>0$ there exists a unique admissible boundary condition for the scalar 
field in $\ads$ leading to a unique AdS-invariant quantization \cite{bf}. In the 
framework of the AdS/CFT correspondence this issue is raised and discussed in 
\cite{bkl, kw,k}. Based on their experience with two-dimensional quantum gravity, 
where the generating functional corresponding to the theory with one branch of 
gravitational dressing is the Legendre transformation of the generating functional 
corresponding to the other branch \cite{k2}, Klebanov and Witten suggested that 
the two different theories are also related by the Legendre transform that interchanges 
the roles of ${\boldsymbol\phi} _0(\Vec{\boldsymbol\gamma})$ and 
${\mathbf A}(\Vec{\boldsymbol\gamma})$ \cite{kw}. Thus the two theories are not 
independent but are, in fact, related to each other by the Legendre transformation. 
They also proposed to interpret ${\mathbf A}(\Vec{\boldsymbol\gamma})$ as the 
expectation value of the conformal operator 
${\boldsymbol{\cal O}}(\Vec{\boldsymbol\gamma})$ namely, 
\begin{equation}\label{oper}
 {\mathbf A}(\Vec{\boldsymbol\gamma})\sim
\langle {\boldsymbol{\cal O}}(\Vec{\boldsymbol\gamma })\rangle\quad.
\end{equation}

It is natural to ask whether the above analogy with two-dimensional quantum gravity
or, equivalently, two-dimensional string theory is deeper. If so, this could lead us 
to a better 
understanding of the AdS/CFT correspondence by our understanding of 2d gravity. In 
the present paper, we will show that 2d gravity methods are indeed appropriate for 
some purposes.

The outline of the paper is as follows. We start in section 2 by describing 
how the world-sheet methods and 2d gravity like scaling arguments 
can be used to describe the $\ads$/CFT correspondence 
in the bosonic case. Our strategy is to start from a world-sheet representation of CFT 
in space-time and, then, to perturb such a theory via marginal 
perturbations of a world-sheet action. The role of these perturbations is to break infinite 
dimensional symmetry in space-time to finite dimensional one. It turns out that 
one of the perturbations is the screening operator of the SL(2) WZW model; i.e. 
the world-sheet action has a form \eqref{sig2}. The latter allows us to interpret 
this case as string on $\ads$ and consider some features of the AdS/CFT correspondence 
in the framework of our formalism. We then go on in section 3 to generalize the results of section 2 for 
the supersymmetric case. This is done by simply adding free world-sheet fermions. 
Finally, section 4 will present our conclusions and directions for future work.

%_______________________      S E C T I O N - 2    _____________________

\section{World-sheet description for $\ads$/CFT correspondence: Bosonic string} 
\renewcommand{\theequation}{2.\arabic{equation}}
\setcounter{equation}{0}

Let us now show how some basic features of the $\ads$/CFT correspondence can be 
caught via world-sheet methods. To do this, we propose a stringy way for 
perturbing 2d CFT's around their critical points. Our strategy is to start with a stringy 
representation of CFT. The latter 
means that there are two Virasoro algebras: one acts on the world-sheet 
and another acts in 
space-time which is two-dimensional (for example, the $\ads$ boundary). Next we 
perturb a world-sheet action by some marginal perturbations such that the 
space-time symmetry becomes finite dimensional. As a result, we obtain a massive FT 
in space-time. Note that a scale for such a theory is introduced by a two-dimensional 
coupling constant or in the framework of the $\ads$/CFT correspondence by 
the radial anti-de-Sitter coordinate, i.e. by $\varphi$. It turns out that the scaling 
argument of David, Distler and Kawai \cite{ddk} applied for 2d gravity is also 
useful to study a scaling limit of this FT \footnote{We refer to \cite{gravrev} for a review 
of two-dimensional gravity.}.

{\it  2.1 More on  the world-sheet description of space-time CFT.} Following 
the picture we 
sketched in subsection 1.2 it seems natural to start with the same set of free 
fields that is used to describe the SL(2) WZW model. Since the Hilbert space of any 
two-dimensional conformal field theory decomposes into holomorphic and 
anti-holomorphic sectors it makes sense to consider one of them, say holomorphic. 
Thus what we have is a free boson $\varphi$ coupled to 
the background charge and a first order bosonic $(\beta,\gamma )$ system of 
weight $(1,0)$. The two-point functions of these fields are normalized as
\begin{equation}\label{norm}
\langle\varphi (z_1)\varphi (z_2)\rangle = -\log \zo
\quad,\quad
\langle\beta(z_1)\gamma(z_2)\rangle = \frac{1}{\zo }
\quad.
\end{equation}
The stress tensor of the free fields coincides with the Sugawara stress tensor of 
the SL(2) WZW model at the level k and is written as 
\begin{equation}\label{sug}
T(z)=\beta\pd\gamma-\oh\pd\varphi\pd\varphi -\frac{1}{\ap }\pd^2\varphi (z)
\quad,
\end{equation}
where $\ap =\sqrt{2({\text k}-2)}$. It is well-known that it provides the world-sheet 
Virasoro algebra with
\begin{equation}\label{vir-w}
{\text L}_n=\oint_{C_0}dz\,z^{n+1}T(z)
\quad, \quad
n\in\mathbf Z
\quad,
\end{equation}
and the central charge $c=\frac{3{\text k}}{{\text k}-2}$. The contour $C_0$ 
surrounds $0$.  

On the other hand, having a pair of the complex space-time coordinates $({\boldsymbol\gamma}, 
\bar{\boldsymbol\gamma} )$, it is straightforward 
to write down the differential operator realization for the space-time Virasoro 
algebras e.g., for the holomorphic sector 
${\mathbf L}_n\sim{\boldsymbol\gamma}^{n+1}
\tfrac{\pd}{\pd{\boldsymbol\gamma}}$. 
At the quantum level the most obvious generalization of that is ${\mathbf L}_n=-
\oint_{C_0}dz\,\gamma^{n+1}\beta (z)$. However, the world-sheet 
reparametrization (conformal) invariance of string theory dictates that the 
vertices should have dimension $1$ in $z$ so that the integral over $dz$ is invariant. 
In other words, the ${\mathbf L}_n$'s have to obey
\begin{equation}\label{vir-vir}
[{\mathbf L}_n,\text L_m]=0
\quad, \quad
n,m\in\mathbf Z
\quad.
\end{equation}
In the framework of the SL(2) WZW model Giveon, Kutasov and Seiberg \cite{gks} 
proposed to modify the formula for $\mathbf L_n$ to 
\begin{equation}\label{vir-s}
{\mathbf L}_n=-\oint_{C_0}dz\,\gamma^{n+1}\beta +\oh\ap (n+1)
\gamma^n\pd\varphi (z)
\quad, \quad
n\in\mathbf Z
\quad.
\end{equation}
Conformal techniques may be used to check that at least for the free fields such 
${\mathbf L}_n$'s indeed generate the Virasoro algebra whose central charge 
is given by ${\mathbf c}=6{\text k}\mathbf k$ with 
${\mathbf k}=\oint_{C_0}dz\,\pd\gamma\gamma^{-1}(z)$. 

What we actually need in practice to do physics in space-time is not only space-time 
symmetry generators defined in terms of world-sheet fields but a world-sheet path 
integral representation for space-time correlators. For example, a possible way to do 
this is
\footnote{Actually, it was assumed in \cite{andre1} that the free actions are perturbed 
by $S_{\text {int}}$. However, because of the space-time conformal invariance, 
$S_{\text {int}}$ has to commute with the ${\mathbf L}_n$'s otherwise the 
Ward identities are broken. So we need to be more careful with perturbations. 
We will return to this point in the next subsection.}
\begin{equation}\label{cor}
\langle \,\,\dots\, \,\rangle_{\text{CFT}}=
\left\lvert \int [d\beta d\gamma ]_p \ep^{-S_0[\beta,\gamma ]}\right\rvert^2
\int [d\varphi] \,\ep^{-S_0[\varphi]}\,\,\dots\quad,
\end{equation}
where $S_0[\beta,\gamma ]$ and $S_0[\varphi]$ are the standard free actions that 
provide the two-point functions \eqref{norm}. $[d\beta d\gamma ]_p $ means that the 
$(\beta,\gamma)$ system has the Bose-sea level $-p$. Note that up to a sign 
$\mathbf k$ coincides with the Bose-sea charge $Q_{\text B}$ of the 
$(\beta ,\gamma)$ system, so $p$ is an eigenvalue of $\mathbf k$ \cite{andre1}.

At this point, it is necessary to make a remark. The ansatz \eqref{cor} means that 
the balance of charges for the free fields is simply
\begin{equation}\label{charge}
\#\beta=\#\gamma
\quad,\quad
\#\bbe=\#\bga
\quad,\quad
\sum\alpha_i=0
\quad.
\end{equation}
Here $\alpha_i$ is given by $\ep^{\alpha_i\varphi}$. This is nothing but the so-called 
Feigin-Fuchs representation used to compute correlators of the SL(2) WZW model. 
The background charge of the field $\varphi$ is compensated by inserting 
the identity operator 
(its conjugate representation $\tilde 1=\exp (-\tfrac{2}{a}\varphi )$) at infinity. 
It is natural to use such an ansatz because we do not know any conjugate 
representation of  the Virasoro generators \eqref{vir-s}. As a simple check, one can 
compute correlators of the ${\mathbf L}_n$'s \cite{andre1}.

Thus, as promised in the introduction, the space-time CFT is described in terms of the 
free world-sheet fields.

{\it  2.2 Perturbations.} So far we have been studying the space-time CFT by the 
world-sheet path integral. It is natural to ask whether the discussion can be extended 
to its perturbations. Since we are interested in the $\ads$/CFT correspondence it is 
reasonable to consider this space-time CFT as the conformal theory which resides on 
the boundary of $\ads$ and try to relate a scale of the perturbed theory with the radial
anti-de-Sitter coordinate. In doing so, what we need is to first realize which perturbations 
lead to $\ads$. In fact this problem is closely connected with the problem of implementing
``space-time boundary conditions'' within the path integral. Indeed, one can start from 
the world-sheet action \eqref{sig2} and try to directly define the boundary CFT. 
We have already discussed this issue in \cite{andre1} where it was proposed to require that 
string vertex operators for the boundary theory be independent of $\varphi$. It is amusing 
that one cannot define the Virasoro and $N=2$ 
generators in a $\varphi$-independent way but one can do so for the $N=4$ 
algebra that is in harmony with the Maldacena conjecture.  However, such a 
construction of the bulk theory remains inconsistent because, while it says how to 
describe the vertex 
operators of physical states, it says nothing about the symmetry breaking. Indeed, one 
of the reasons why $\ads$ is special is that in this case the boundary CFT is 
two-dimensional, so the corresponding conformal symmetry is infinite dimensional. 
On the other hand, string theory on $\ads$ has $\text{SL(2)}\times\text{SL(2)}$ 
symmetry which corresponds to the global part of the two-dimensional conformal 
symmetry. After this is noted, it immediately comes to mind to realize this 
symmetry breaking by the perturbations. It is natural to ask whether all such perturbations 
results in $\ads$. It turns out that this is not the case, so we have to be careful at this point.

Following the ideas we sketched, modify our ansatz \eqref{cor} to 
\begin{equation}\label{cor-p}
\langle \,\,\dots\, \,\rangle =
\langle \,\,\ep^{S_{\text{int}} }
\,\dots\, \,\rangle_{\text{CFT}}=
\left\lvert \int [d\beta d\gamma ]_p \ep^{-S_0[\beta,\gamma ]}\right\rvert^2
\int [d\varphi] \,\ep^{-S_0[\varphi]+S_{\text{int}} }\,\,\dots
\quad.
\end{equation}
So what we have done was to perturb the free actions by 
$S_{\text{int}}[\varphi ;\beta,\gamma;\bga ,\bbe ]$. By analogy with the SL(2) WZW 
model, we require 
\begin{equation}\label{w-s}
[S_{\text{int}},\text L_n]=[S_{\text{int}},\bar{\text L}_n]=0
\quad,\quad
n\in\mathbf Z
\quad.
\end{equation}
In other words, $S_{\text{int}}$ is marginal for the world-sheet theory.  This is, 
however, not the hole story. A novelty is to require
\begin{equation}\label{s-t}
[S_{\text{int}},{\mathbf L}_n]=[S_{\text{int}},\bar{\mathbf L}_n]=0
\end{equation}
only for $n=0, \pm 1$. What this means in practice is as follows. We have no longer 
the conformal theory 
in space-time. To be more precise, what we got is the so-called quasi-conformal 
theory \cite{bpz}. Even though we have broken the infinite dimensional symmetry to 
its global part $\text{SL(2)}\times\text{SL(2)}$, it is instructive to ask whether 
we got string theory on $\ads$. However, before answering this question, let us 
give a few examples of $S_{\text{int}}$ obeying \eqref{w-s}-\eqref{s-t}.

(1) Following the ideas sketched in subsection 1.2, it is natural to try the screening 
operator of the SL(2) WZW model . For simplicity, let us restrict ourselves to its 
holomorphic part, namely, $S_+=\oint dz\,\beta\exp (-\tfrac{2}{a}\varphi )$. Then, 
a simple algebra shows that it indeed obeys \eqref{w-s}-\eqref{s-t}. The latter means 
that the operator 
\begin{equation}\label{cos}
{\cal O}_0=\int d^2z\,\beta\bbe\,\ep^{-\frac{2}{a}\varphi} (z,\bz)
\end{equation}
can be used as $S_{\text{int}}$.

(2) We now want to reconsider the above derivation. The idea is to 
modify $S_+$ to $\oint dz\,\beta^x\exp (y\varphi )$. The analysis proceeds as above 
and leads to a new solution $S_-=\oint dz\,\beta^{{\text k}-2}\exp (-a\varphi )$. 
Thus we can use 
\begin{equation}\label{cos2}
{\cal O}_1=\int d^2z\,(\beta\bbe)^{{\text k}-2}\ep^{-a\varphi} (z,\bz)
\end{equation}
as the perturbation of the free theory. It is interesting to note that $S_-$ is nothing 
but the second screening operator of the SL(2) WZW model found by Dotsenko \cite{d}. 

(3) Alternatively, it is possible to find a solution of the constraints 
\eqref{w-s}-\eqref{s-t} in a pure algebraic way. In doing so, one has to keep in mind 
that the ${\text L}_n$'s and ${\mathbf L}_m$'s commute. It automatically follows, then, 
that any polynomial in the ${\mathbf L}_m$'s  is a solution of \eqref{w-s}. Thus the 
problem is reduced to finding polynomials that 
obey \eqref{s-t}. It is straightforward to write down a solution of the problem. 
It is simply given by the quadratic Casimir operator of sl(2) namely,
\begin{equation}\label{cos3}
{\mathbf C}_2=-{\mathbf L}_0^2+\oh({\mathbf L}_1{\mathbf L}_{-1}+
{\mathbf L}_{-1}{\mathbf L}_1 )
\quad.
\end{equation}
This means that a formal expression 
${\cal O}_2={\mathbf C}_2\bar{\mathbf C}_2$ may be used as the 
perturbation $S_{\text{int}}$. However, it is not clear whether such an expression may 
be written in a local form like $\int d^2z\, V(z,\bz )$. 

Now let us turn to a geometrical interpretation of the above results. Heuristically, 
the idea is to integrate away the auxiliary $\beta ,\bbe$ fields in the world-sheet path 
integral in order to get a term in the sigma model action that corresponds to 
$G_{\mu\nu}\pd X^\mu\pd X^\nu$. The last would allow us to reconstruct 
the space-time geometry. It is well known that it works fine for ${\cal O}_0$ where 
one easily finds the $\ads$ metric. 
But it fails for ${\cal O}_1$ and ${\cal O}_2$. The problem is that we have no longer 
 a liner dependence on $\beta\bbe $ in $S_{\text{int}}$. So it is desirable to have new 
ways of understanding string theory on curved spaces to find the geometrical 
interpretation for these perturbations.

{\it  2.3 More on string theory on} $\ads$. As we have seen above, string theory 
on $\ads$ can be 
described by perturbing the free field actions within the world-sheet path integral. 
The result of such a perturbation is breaking the infinite dimensional conformal 
symmetry in space-time. From a local observer point of view who lives on the 
$(\boldsymbol\gamma, \bar{\boldsymbol\gamma} )$ plane, what we got is a 
two-dimensional Field Theory in a vicinity of its critical point. A simple idea behind 
some of the recent advances in studying the AdS/CFT correspondence 
is that the anti-de-Sitter coordinate $\boldsymbol\varphi $ is responsible for a 
scale in this Field Theory \cite{adsrev}. We will now attempt to make more precise 
this statement in the case of $\ads$. Moreover, we will show that it is in harmony 
with what we proposed in the previous subsection.

Our ansatz for the Field Theory partition function is that 
\begin{equation}\label{pf-r}
Z[{\mathbf r}]=\langle \,
\delta\bigl(\int d^2z\,\beta\bbe\,\ep^{-\frac{2}{a}\varphi} (z,\bz)-{\mathbf r}^{-1}
\bigr)\, \rangle_{\text{CFT}}
\quad,
\end{equation}
where $0\leq{\mathbf r}\leq\infty$.

As a preliminary check, note that the Laplace transformed partition function 
\begin{equation}\label{pf-c}
Z[t_0]={\cal L}Z[{\mathbf r}]=
\int_0^\infty d{\mathbf r}\,\ep^{-t_0/{\mathbf r}}Z[{\mathbf r}]
\end{equation}
has the expected form \eqref{cor-p} with the perturbation 
$S_{\text{int}}=-t_0\int d^2z\,\beta\bbe\,\ep^{-\frac{2}{a}\varphi}$. 

Now we will recall how physical states (primary conformal operators) 
appear in the space-time CFT. Several different 
proposals are known to define them. One is based on the idea that the $\mathbf L_n$'s 
defined in \eqref{vir-s} are in fact the coefficients of the Laurent expansion
\begin{equation}\label{vir1}
{\mathbf L}_n=\oint_{C_0}d{\boldsymbol\gamma}\,{\boldsymbol\gamma}^{n+1}
\mathbf T({\boldsymbol\gamma})
\quad,
\end{equation}
where $\mathbf T({\boldsymbol\gamma})$ is the stress tensor of the boundary theory. 
The realization of the primary fields (their holomorphic parts) via this definition is 
then \footnote{Alternatively, one can think about the ${\mathbf L}_n$'s as 
${\mathbf L}_n{\boldsymbol\Phi}^\Delta ({\boldsymbol\gamma})=
\oint_{C_{\boldsymbol\gamma}}d{\boldsymbol\gamma}^\prime\,
({\boldsymbol\gamma}^\prime -{\boldsymbol\gamma})^{n+1}
\mathbf T({\boldsymbol\gamma}^\prime )
{\boldsymbol\Phi}^\Delta ({\boldsymbol\gamma})$. See \cite{andre2} for a 
discussion of this approach within string theory on $\ads$.}
\begin{equation}\label{pf}
[{\mathbf L}_n,{\boldsymbol\Phi}_m^\Delta]=
\bigl(n(\Delta-1)-m\bigr){\boldsymbol\Phi}_{n+m}^\Delta
\quad.
\end{equation}
Here the ${\boldsymbol\Phi}_m^\Delta$'s are the Laurent modes of the primary 
field whose conformal weight (dimension) is $\Delta$ i.e., 
${\boldsymbol\Phi}^\Delta ({\boldsymbol\gamma})=
\sum_n {\boldsymbol\gamma}^{-n-\Delta}{\boldsymbol\Phi}_m^\Delta$. An 
example of such a realization within string theory on $\ads$ times some compact 
manifold X was proposed in \cite{gks}. Giveon and co-workers  simply form a 
vertex operator for 
${\boldsymbol\Phi}_m^\Delta$ by dressing a spinless operator $V(z,\bz )$ of CFT on 
X by the SL(2) WZW primary field. Explicitly,  
\begin{equation}\label{pf-gks}
{\boldsymbol\Phi}_{m\,\bar m}^\Delta=\int d^2z\,
{\boldsymbol\gamma}^{j+m}\bar{\boldsymbol\gamma}^{j+\bar m}
\ep^{\frac{2}{a}j\varphi }\,V(z,\bz )
\quad.
\end{equation}
Then it follows from Eq. \eqref{pf} that for the primary field 
${\boldsymbol\Phi}^\Delta$ the conformal weight is given by $\Delta =j+1$.

Now we are ready to compare our ansatz \eqref{pf-r} to ``experiment''. First, let us to 
define the scaling dimension $\Delta ({\boldsymbol\Phi})$ of the physical operator as
it is done within 2d gravity \cite{ddk} namely,

\begin{equation}\label{scal-d}
\frac{1}{Z[{\mathbf r}]}\langle \,\delta
\bigl(\int d^2z\,\beta\bbe\,\ep^{-\frac{2}{a}\varphi} (z,\bz)-{\mathbf r}^{-1}\bigr)
{\boldsymbol\Phi}
\, \rangle_{\text{CFT}}\sim {\mathbf r}^{-1+\Delta ({\boldsymbol\Phi})}
\quad, \quad {\mathbf r}\rightarrow 0
\quad.
\end{equation}
We can use a simple scaling argument to evaluate the one-point functions of the 
operators \eqref{pf-gks}. Indeed, shifting the zero mode of the field $\varphi$
\begin{equation}\label{shift}
\varphi\rightarrow\varphi +\frac{a}{2}\log {\mathbf r}
\end{equation}
and assuming that the path integral measure $[d\varphi] $ is invariant under 
such a shift we find that the scaling dimension is given by 
$\Delta ({\boldsymbol\Phi})=j+1$. 

Moreover, taking into account the contribution from the $\delta$-function we get 
for the non-normalizable one-point functions
\begin{equation}\label{ass-w}
\langle \,\delta
\bigl(\int d^2z\,\beta\bbe\,\ep^{-\frac{2}{a}\varphi} (z,\bz)-{\mathbf r}^{-1}\bigr)
{\boldsymbol\Phi}^\Delta_{m\,\bar m}
\, \rangle_{\text{CFT}}\sim {\mathbf r}^{\Delta ({\boldsymbol\Phi})}
\quad.
\end{equation}

Now we can easily read off some interesting conclusions. One of the first observations is 
that the conformal weights (dimensions with respect to ${\mathbf L}_0$) 
of the operators \eqref{pf-gks} coincide with their scaling dimensions namely, 
$\Delta=\Delta( {\boldsymbol\Phi})$. This gives us a 
hint that one should try to catch the dynamics of such operators via the effective 
action for scalar fields in the $\ads$ background. Further evidence in favor of this 
suggestion is provided by comparing Eq. \eqref{ass} to Eq. \eqref{ass-w}. Indeed, we can 
interpret the last equation as the asymptotic behaviour of the effective scalar field, 
namely, its part related to physical fluctuations. The latter assumes that 
${\mathbf A}(\Vec{\boldsymbol\gamma})$ has the following representation:
\begin{equation}\label{kw}
{\mathbf A}(\Vec{\boldsymbol\gamma})\sim{\mathbf r}^{-\Delta}
\langle \,\delta
\bigl(\int d^2z\,\beta\bbe\,\ep^{-\frac{2}{a}\varphi} (z,\bz)-{\mathbf r}^{-1}\bigr)
{\boldsymbol\Phi}^\Delta (\Vec{\boldsymbol\gamma})
\, \rangle_{\text{CFT}}
\quad, \quad 
{\mathbf r}\rightarrow 0
\quad,
\end{equation}
where ${\boldsymbol\Phi}^\Delta (\Vec{\boldsymbol\gamma})=\sum_{m\, \bar m}
{\boldsymbol\gamma}^{-m-\Delta}\bar{\boldsymbol\gamma}^{-\bar m-\Delta}
{\boldsymbol\Phi}_{m\,\bar m}^\Delta$.
This is exactly what Klebanov and Witten proposed in \cite{kw} (see \eqref{oper}).

Next, let us go on to look more carefully  at the picture we propose. According to our 
discussion, we have the same set of the physical operators \eqref{pf-gks} for both the CFT 
and string theory on $\ads$. The only difference between these theories is the 
ansatz used to compute space-time correlators. Thus we have the explicit example 
of the translation of physical states of one theory to another, i.e. the example of the 
't Hooft Holographic principle \cite{h,suss}.

{\it  2.4 Towards RG analysis.} Up to now our discussion has not been sensitive to 
the second 
term in the asymptotic behaviour \eqref {ass} that is responsible for a ``source''. A 
question remains, however, as to what is a good world-sheet approximation for it. 
A possible answer to this question immediately comes to mind just by looking at the 
ansatz \eqref{corr} for the AdS/CFT correspondence.  So we would like to relate
world-sheet couplings $t_i$ with the ``sources''. 

As a preliminary check, let us consider the perturbation 
$t_0\int d^2z\,\beta\bbe\,\ep^{-\frac{2}{a}\varphi} $. 
As before, we can shift $\varphi$ to determine the scaling dimension of the operator 
$\int d^2z\,\beta\bbe\,\ep^{-\frac{2}{a}\varphi} $. Thus $\Delta_0 =0$. On the other 
hand, a scaling argument shows that $Z[{\mathbf r}]\sim {\mathbf r}$ as 
${\mathbf r}\rightarrow 0$. Using the Laplace transform \eqref{pf-c}, we find that 
the partition function $Z[t_0]$ scales as $t_0^2$. So for $\langle\,t_0\,\rangle$, we 
have $t_0^3$ \footnote{$\langle\,1\,\rangle$ corresponds to the 
partition function $Z$ of the perturbed theory. So $\langle\,t_0\,\rangle$ is 
simply $t_0Z$.  }. In terms of 
$\mathbf r$ it corresponds to ${\mathbf r}^2$. This is 
the expected form of the asymptotic behavior for zero conformal dimension. 

We want now to repeat this analysis in a general case. As before, the first step is to find 
the scaling law for $\langle\,t_i\,\rangle$ in terms of $t_0$. In fact, the ansatz 
\eqref{corr} assumes that all $\langle\,t_i{\cal O}_i\,\rangle $ should scale in the 
same way, i.e. like $t_0^2$. The latter means that $t_i$ behaves as $t_0^{1-\Delta_i}$.
So what we find for $\langle\,t_i\,\rangle$ is then $t_0^{3-\Delta_i}$. In terms of 
$\mathbf r$ it is replaced by ${\mathbf r}^{2-\Delta_i}$. This is exactly what we need to 
push our interpretation of the world-sheet coupling constants as the space-time 
``sources'', namely, in the scaling limit 
\begin{equation}\label{coup}
{{\boldsymbol\phi }_0}_i(\Vec{\boldsymbol\gamma})\sim
{\mathbf r}^{\Delta_i-2}{\cal L}^{-1}\langle\,t_i\,\rangle
\quad, \quad 
{\mathbf r}\rightarrow 0
\quad.
\end{equation}
We would like to make a few comments:

1. Although the representations for $\mathbf A$ and ${\boldsymbol\phi} _0$ given by 
\eqref{kw}-\eqref{coup} look in many ways attractive, we have to stress their 
speculative character. They rest on the scaling argument only, and so further 
work is needed to prove them strictly. Moreover, we do not know any similar
representation for the product 
${\boldsymbol \phi} _0{\boldsymbol{\cal O}}(\Vec{\boldsymbol\gamma})$ in Eq. 
\eqref{corr}. What we can only refer to the examples of subsection 2.2.

2. As we saw in the previous subsection, the scale of the perturbed theory is provided 
by the anti-de-Sitter coordinate 
$\varphi$. This means that the coupling constant (source) in Eq. \eqref{corr} becomes 
running, i.e. ${\boldsymbol\phi} _0(\Vec{\boldsymbol\gamma}, {\mathbf r})$. The 
latter allows to define the corresponding $\beta$-function as 
\begin{equation}\label{beta}
\beta_i={\mathbf r}\pd_{\mathbf r}
{{\boldsymbol\phi} _0}_i(\Vec{\boldsymbol\gamma}, {\mathbf r})
\quad.
\end{equation}
It is natural to suggest that the scaling limit of the running coupling is described via 
${\cal L}^{-1}\langle\,t_i\,\rangle$. Thus the linearized $\beta$-function is simply 
\begin{equation}\label{beta-l}
\beta_i=(2-\Delta_i){{\boldsymbol\phi} _0}_i(\Vec{\boldsymbol\gamma}, {\mathbf r})
\quad.
\end{equation}
A simple observation is, then, that $\beta_i$ vanishes for $\Delta_i=2$, i.e. for the 
operator whose dimension equals $2$. This gives us a hint that one should catch 
dynamics of such the operator via the effective action for the massless scalar field. This is 
in harmony with the formula \eqref{kpz}. 

3. Note that the situation that we are considering here is more subtle than the one in 
two-dimensional gravity in which the Legendre transformation of the generating 
functional corresponds to the other branch of gravitational dressing. The crucial point is 
that in the problem at hand a transformation has to know about the unitary bound. 
In other words, it should be defined only for the range $0\leq\Delta\leq 2$. So a naive 
attempt to adopt the 2d gravity analysis \cite{k2} fails.

%_______________________      S E C T I O N - 3   _____________________

\section{Inclusion of Supersymmetry} 
\renewcommand{\theequation}{3.\arabic{equation}}
\setcounter{equation}{0}

According to the Maldacena conjecture, in the situation that we are considering the 
theory on the boundary of $\ads$ has to possess the $N=4$ superconformal algebra as 
the symmetry algebra. Thus, in this section we will generalize and apply the previous results to the 
supersymmetric case.

{\it  3.1 World-sheet description of space-time SCFT}. Following the original analysis of \cite{gks}, 
the world-sheet description of $N=4$ SCFT on the boundary of $\ads$ is as follows. 

The $\ads$-part (its holomorphic sector) is described by the same set of the free fields we 
used in subsection 2.1 namely, $(\varphi ,\beta ,\gamma )$. The only difference is 
that the level k is shifted as ${\text k}\rightarrow {\text k}+2$. The latter 
is due to free fermions with the two-point functions
\begin{equation}\label{fer}
\langle\psi^i(z_1)\psi^j(z_2)\rangle = \frac{\eta^{ij}}{\zo }
\quad,
\end{equation}
where $i,j=0,\pm;\,\,\eta^{00}=-1,\,\eta^{+-}=\eta^{-+}=2$. It is well known that 
such 
fermions are needed to define the world-sheet $N=1$ superconformal algebra 
that is standard within the Neveu-Schwarz-Ramond (NSR) formulation of superstring theories.

The ${\mathbf S}^3$-part (its holomorphic sector) can be described in a similar way. 
So we introduce a free boson $\varphi_c$ coupled to a background charge, the 
first order bosonic $(\beta_c,\gamma_c)$ system of weight $(1,0)$ and three fermionic 
fields $\psi^i_c$. The two-point functions of these fields are normalized as 
\begin{equation}\label{norm-c}
\langle\varphi_c (z_1)\varphi_c (z_2)\rangle = -\log \zo
\quad,\quad
\langle\beta_c(z_1)\gamma_c(z_2)\rangle = \frac{1}{\zo }
\quad,\quad
\langle\psi^i_c(z_1)\psi^j_c(z_2)\rangle = \frac{\eta_c^{ij}}{\zo }
\quad,
\end{equation}
where $i,j=0,\pm ;\,\,\eta_c^{00}=1,\,\eta_c^{+-}=\eta_c^{-+}=2$.

The stress tensor of the bosonic fields is given by
\begin{equation}\label{sug-c}
T_c(z)=\beta_c\pd\gamma_c-\oh\pd\varphi_c\pd\varphi_c 
+\frac{i}{\ap }\pd^2\varphi_c (z)
\quad,
\end{equation}
where $\ap =\sqrt{2{\text k}}$. It coincides with the Sugawara stress tensor of the 
SU(2) WZW model at the level ${\text k}-2$ 
\begin{equation}\label{sug-cc}
T_c(z)=\frac{1}{{\text k}}\eta_{c\,ij}\,J_c^iJ_c^j(z)
\quad,
\end{equation}
such that the currents are 
\begin{equation}\label{wak}
J^-_c(z)=-i\beta_c (z)
\quad, \quad
J^0_c(z)=\beta_c\gamma_c+\frac{i}{2}a\pd\varphi_c (z)
\quad, \quad
J^+_c(z)=-i\beta_c\gamma_c^2+a\gamma_c\pd\varphi_c+
i({\text k}-2)\pd\gamma_c(z)
\quad.
\end{equation} 
In a similar way, the stress tensor of the free fermions coincides with the Sugawara 
stress tensor of the SU(2) WZW model at the level $2$. The corresponding currents 
are given by
\begin{equation}\label{fc}
j^-_c(z)=i\psi^-_c\psi^0_c (z)
\quad,\quad
j^0_c(z)=\oh\psi^+_c\psi^-_c (z)
\quad,\quad
j^+_c(z)=i\psi^+_c\psi^0_c (z)
\quad.
\end{equation}

As to a manifold X, it is usually associated with ${\mathbf T}^4$ or ${\mathbf K3}$. The 
explicit choice is not crucial for what follows; however, to be more precise let us take 
the four torus as X. Then, the ${\mathbf T}^4$-part (its holomorphic sector) can be 
described by four scalar fields $X^\mu$ without background charges together with 
their fermionic partners $\chi^\mu$, $\mu=1,\dots ,4$. Their two-point functions 
can be normalized as it was done in \eqref{norm-c}. Finally, one should keep in mind the 
corresponding superconformal ghosts that are needed to cancel the conformal 
anomaly \cite{gks, fms}.

Having the above set of the free fields, we now wish to realize the superconformal algebra 
of the boundary SCFT. This can be done as in \cite{gks}. Let us restrict ourselves to the 
holomorphic sector. The Virasoro generators \eqref{vir-s} are modified to 
\begin{equation}\label{vir-4}
{\mathbf L}_n=-\oh\oint_{C_0}dz\,\gamma^n
\Bigl( 2\gamma\beta +\ap (n+1)\pd\varphi -(n^2-1)\psi^+\psi^-+
i(n^2+n)\gamma^{-1}\psi^+\psi^0+
i(n^2-n)\gamma\psi^-\psi^0 
\Bigr)(z)
\end{equation}
where $a$ is now $\sqrt {2{\text k}}$.

The SU(2) generators are provided by the ${\mathbf S}^3$-part or, equivalently, by the 
supersymmetric SU(2) WZW model. Explicitly,
\begin{equation}\label{color-4}
{\mathbf T}^i_n=\oint_{C_0}dz\, \gamma^n \bigl (\,J^i_c  +j^i_c \,\bigr)(z)
\quad, \quad
n\in\mathbf Z
\quad.
\end{equation}
A simple algebra shows that such defined ${\mathbf L}_n$'s and ${\mathbf T}^a_n$'s
obey the commutation relations of the $N=4$ superconformal algebra whose central 
charge is $\mathbf c=6{\text k}\mathbf k$. 

It is known that one of the drawbacks of the NSR formalism is that it is not manifestly 
supersymmetric. In the problem at hand it is rather difficult to explicitly write down 
all fermionic generators. For simplicity, we will restrict ourselves to the global 
generators. In the NSR sector they are simply \cite{gks}
\begin{equation}\label{}
{\mathbf Q}_{\pm\oh }^{\alpha}=\oint_{C_0}dz\, \ep^{-\oh\phi}S^{\alpha} (z)
\quad,\quad
\alpha=\pm,{\dot\pm}
\quad.
\end{equation}
Here $\phi$ is the scalar field which appeared by the bosonization of the superconformal 
ghosts. $S^{\alpha}$ are the so-called spin fields of the $N=1$ 
world-sheet superconformal algebra. Note that the spin fields are built only via the 
fermions $\psi,\psi_c,\chi$.

Finally, let us generalize our ansatz \eqref{cor}. For example, this can be done as
\begin{equation}\label{cor-s}
\langle \,\,\dots\, \,\rangle_{\text{SCFT}}=
\left\lvert \int [d\beta d\gamma ]_p \ep^{-S_0[\beta,\gamma ]}\right\rvert^2
\int [d\varphi] \,\ep^{-S_0[\varphi]}
\left\lvert \int [d\beta_cd\gamma_c ]_0 \ep^{-S_0[\beta_c,\gamma_c ]}
\right\rvert^2
\int [d\varphi_c] \,\ep^{-S_0[\varphi_c]}
\,\,\dots\quad,
\end{equation}
Above, we have omitted the additional scalars fields as well as the fermions and the 
superconformal ghosts. The path integral measures for these fields are 
standard (see e.g., \cite{fms}). We use the Feigin-Fuchs representation again, so 
we require the following balance of charges 
\begin{equation}\label{charge-s}
\#\beta=\#\gamma
\quad,\quad
\#\bbe=\#\bga
\quad,\quad
\sum\alpha_i=0
\quad,\quad
\#\beta_c=\#\gamma_c
\quad,\quad
\#\bbe_c=\#\bga_c
\quad,\quad
\sum{\alpha_c}_i=0
\quad,
\end{equation}
where $\alpha_i$ and ${\alpha_c}_i$ are given by $\ep^{\alpha_i\varphi}$ and 
$\ep^{i{\alpha_c}_i\varphi_c}$, respectively. It should be noted that unlike the 
$(\beta,\gamma)$ system of the $\ads$-part, the $(\beta_c,\gamma_c)$ 
system of the ${\mathbf S}^3$-part has a zero Bose sea level.

{\it  3.2 The world-sheet description of superstring theory on} 
$\ads\times {\mathbf S}^3\times \text X$. To get superstring theory from SCFT, 
we follow the same strategy as we proposed in section 2. So we perturb the free 
world-sheet actions by marginal perturbations such that the space-time 
symmetry becomes finite dimensional while the $G_{\mu\nu}\pd X^\mu\pd X^\nu$ term with 
the $\ads$ metric appears in the world-sheet action. As a result, we get a supersymmetric FT in 
space-time. Again, a scale for this theory is provided by a world-sheet coupling 
constant or, equivalently, by the radial anti-de-Sitter coordinate $\varphi$. 

Let us first modify the ansatz \eqref{cor-s} to 
\begin{equation}\label{corb-s}
\langle \,\,\dots\, \,\rangle =
\langle \,\,\ep^{\hat S_{\text{int}} }
\,\dots\, \,\rangle_{\text{SCFT}}
\quad.
\end{equation}
We now require that $\hat S_{\text{int}}$ obeys \footnote{Strictly speaking, we also need 
to require that $[\hat S_{\text{int}},\hat{\text G}_r]=
[\hat S_{\text{int}},\bar{\hat{\text G}}_r]=0$, where the 
$\hat{\text G}_r(\bar{\hat{\text G}}_r)$'s are the $N=1$ fermionic generators. 
However, it proves irrelevant for what follows.}
\begin{equation}\label{w-ss}
[\hat S_{\text{int}},\hat{\text L}_n]=[\hat S_{\text{int}},\bar{\hat{\text L}}_n]=0
\quad,\quad
n\in\mathbf Z
\quad,
\end{equation}
where $\hat{\text L}_n$ means a total world-sheet Virasoro generator. So 
$\hat S_{\text{int}}$ is marginal within the world-sheet theory.

To extend our bosonic analysis to the supersymmetric case, we need, in addition to 
\begin{equation}\label{ss-t}
[\hat S_{\text{int}},{\mathbf L}_n]=[\hat S_{\text{int}},\bar{\mathbf L}_n]=0
\quad,\quad\text{only for}\quad n=0,\pm 1
\quad,
\end{equation}
that $\hat S_{\text{int}}$ obeys 
\begin{equation}\label{ss-t2}
[\hat S_{\text{int}},{\mathbf T}_n^i]=[\hat S_{\text{int}},\bar{\mathbf T}_n^i]=0
\quad,\quad\text{for}\quad n=0
\quad\text{and}\quad
[\hat S_{\text{int}},{\mathbf Q}_r^\alpha]
=[\hat S_{\text{int}},\bar{\mathbf Q}_r^\alpha]=0
\quad,\quad\text{for}\quad r=\pm\oh
\quad.
\end{equation}
Thus we have no longer a superconformal theory in space-time. To be more 
precise, what we got is the global $N=4\times N=4$ algebra. In the above, we restrict 
ourselves to the Neveu-Schwarz sector. However, the generalization to others is 
straightforward. 

Let us now give a  few examples of $\hat S_{\text{int}}$ obeying \eqref{w-ss}
-\eqref{ss-t2}.

(1) As in section 2, let us try the screening operator of the SL(2) WZW model. A precisely 
analogous computation shows that it indeed obeys \eqref{w-ss}-\eqref{ss-t2}. Thus 
the operator 
\begin{equation}\label{coss}
\hat{\cal O}_0=\int d^2z\,\beta\bbe\,\ep^{-\frac{2}{a}\varphi} (z,\bz)
\end{equation}
can be used as $\hat S_{\text{int}}$.

There is a point we should mention. Integrating the auxiliary fields $\beta,\bbe$ in the 
world-sheet path integral away, we get a non-linear term in the sigma model action 
which corresponds to $G_{\mu\nu}\pd X^\mu\pd X^\nu$ with the metric of 
$\ads$ while the fermionic terms remains quadratic. Here an analogy with the Green-Schwarz (GS) 
formulation appears because such behavior reminds us of results of gauge fixing the 
GS action where the fermionic term becomes quadratic (see, e.g., \cite{kt,pes}).

(2) It is also not difficult to find one more operator obeying the above constraints. A 
simple algebra shows that it is given by 
\begin{equation}\label{coss2}
\hat{\cal O}_1=\int d^2z\,(\beta\bbe)^{\text k}\ep^{-a\varphi} (z,\bz)
\quad.
\end{equation}

(3) Obviously, the quadratic Casimir operator of the global $N=4\times N=4$ algebra,
\begin{equation}\label{coss3}
\hat{\mathbf C}_2=-\mathbf L_0^2+\oh(\mathbf L_1\mathbf L_{-1}+
\mathbf L_{-1}\mathbf L_1 )+\eta_{c\,ij}\,{\mathbf T}^i_0{\mathbf T}^j_0+
2\epsilon_{\alpha\beta}{\mathbf Q}^\alpha_{(\oh}{\mathbf Q}^\beta_{-\oh)}
\end{equation}
obeys the constraints. However, it is not clear how to rewrite 
$\hat{\text C}_2\bar{\hat{\text C}}_2$ in a local form.

It is interesting to note that the screening operators of the SU(2) WZW model 
$\int d^2z\,\beta_c\bbe_c\,\ep^{\frac{2i}{a}\varphi_c}$ and 
$\int d^2z\,(\beta_c\bbe_c)^{-\text k}\ep^{-ia\varphi_c}$ are marginal for both 
space-time and world-sheet theories.

We conclude this subsection with a brief discussion of $\hat{\cal O}_1$. On the one hand, 
it is the second screening operator of the SL(2) WZW model. Moreover, it is known that 
it becomes the second screening operator for the Liouville theory under the 
Drinfeld-Sokolov reduction. The latter is similar to a light-cone-like gauge in the 
context of the AdS/CFT correspondence \cite{adsrev}. On the other hand, the second 
screening operator is crucial 
for a strong coupling regime of 2d gravity \footnote{ The interested reader is referred to 
lectures of Gervais \cite{ger}.}.  So putting the two facts together, we suggest that the 
correct perturbation of the free actions is given by
\begin{equation}\label{str}
\hat S_{\text{int}}=\hat{\cal O}_0+\hat{\cal O}_1
\quad.
\end{equation}
Let us give one more piece of evidence in favor of this suggestion. Obviously, these operators 
coincide for ${\text k}=1$. In this sense, something should happen as one approaches 
this value. This problem was first discussed in \cite{ds} and later in \cite{sw}. However, 
our description is different because in fact it assumes an analogy with the 
so-called $c=1$ barrier in two-dimensional quantum  gravity \cite{gravrev}.

{\it  3.3 More comments on superstring theory}. The extension of our analysis of subsections 
2.3 and 2.4 to the supersymmetric case is straightforward. Therefore, we only 
summarize the most relevant formulae.

A new ansatz for the partition function is given by 
\begin{equation}\label{spf-r}
\hat Z[{\mathbf r}]=\langle \,
\delta\bigl(\int d^2z\,\beta\bbe\,\ep^{-\frac{2}{a}\varphi} (z,\bz)-{\mathbf r}^{-1}
\bigr)\, \rangle_{\text{SCFT}}
\quad.
\end{equation}
Of course, one can easily get the expected form \eqref{corb-s} with 
$\hat S_{\text{int}}=-t_0\int d^2z\,\beta\bbe\,\ep^{-\frac{2}{a}\varphi} (z,\bz)$ by 
the Laplace transformation.

It seems natural to define the scaling dimension 
$\Delta (\hat{\boldsymbol\Phi})$ of the physical operator as 
\begin{equation}\label{sscal-d}
\frac{1}{\hat Z [{\mathbf r}]}\langle \,\delta
\bigl(\int d^2z\,\beta\bbe\,\ep^{-\frac{2}{a}\varphi} (z,\bz)-{\mathbf r}^{-1}\bigr)
\hat{\boldsymbol\Phi}
\, \rangle_{\text{SCFT}}\sim {\mathbf r}^{-1+\Delta (\hat{\boldsymbol\Phi})}
\quad, \quad {\mathbf r}\rightarrow 0
\quad,
\end{equation}
and, then, to probe the vertex operators proposed by Giveon and co-workers \cite{gks}. 
In the supersymmetric case these operators look like hatted versions of \eqref{pf-gks}. 
On the one hand, the commutation relations with the Virasoro generators 
show that their conformal dimension is given by $\Delta=j+1$. On the other hand, 
the scaling argument gives the scaling dimension as 
$\Delta (\hat{\boldsymbol\Phi})=j+1$. Therefore, as in subsection 2.3, we propose 
the following representation of  ${\mathbf A}$       
\begin{equation}\label{skw}
{\mathbf A}(\Vec{\boldsymbol\gamma})\sim{\mathbf r}^{-\Delta}
\langle \,\delta
\bigl(\int d^2z\,\beta\bbe\,\ep^{-\frac{2}{a}\varphi} (z,\bz)-{\mathbf r}^{-1}\bigr)
\hat{\boldsymbol\Phi}^\Delta (\Vec{\boldsymbol\gamma})
\, \rangle_{\text{SCFT}}
\quad, \quad 
{\mathbf r}\rightarrow 0
\quad,
\end{equation}
with $\hat{\boldsymbol\Phi}^\Delta (\Vec{\boldsymbol\gamma})=\sum_{m\, \bar m}
{\boldsymbol\gamma}^{-m-\Delta}\bar{\boldsymbol\gamma}^{-\bar m-\Delta}
\hat{\boldsymbol\Phi}_{m\,\bar m}^\Delta$. This is the world-sheet representation 
of the Klebanov-Witten proposal \cite{kw}. 

As to the space-time ``sources'' ${{\boldsymbol\phi} _0}_i$, we again propose that their 
scaling limit is described via ${\cal L}^{-1}\langle\,t_i\,\rangle$. This allows us to find 
the linearized $\beta$-function. Explicitly, 
$\beta_i=(2-\Delta_i){{\boldsymbol\phi} _0}_i$.

We here conclude this section with some speculations about Eq.\eqref{kpz} which gives 
the relation between the mass $\mathbf m$ of the effective scalar field in $\ads$ and the 
conformal dimension $\Delta$ of the corresponding boundary operator.

(1) In general, it is not clear how to define the S-matrix within Field Theory in 
AdS spaces. As a result, we cannot consistently define masses as poles in the scattering 
amplitudes. So the best that we can do with the problem at hand is to take the flat limit 
$l\rightarrow\infty$. The latter is equivalent to the limit ${\text k}\rightarrow\infty$. 
In free fields terms, it means the following rescaling for the scalar field 
$\varphi\rightarrow\sqrt{\frac{2}{\text k}}\varphi$. Thus the exponent in 
Eq.\eqref{pf-gks} becomes $\ep^{j\varphi}$. As to the fields $\gamma$ and $\bga$, it is 
useful to bosonize them in the standard way, for instance, 
$\gamma=\ep^{i\sigma^1-\sigma^2}$. Now we can define the mass just as 
the sum of exponents. So we find ${\mathbf m}^2=j^2$. Note that the bosonization 
makes clear that $\gamma$ and $\bga$ do not contribute. However, there is a 
contribution of other fields. Since we are only interested in a dependence of $j$ we 
simply modify the formula to ${\mathbf m}^2=j^2+const$. There are two facts that 
we should recall. The first is that the conformal dimension of the operators 
\eqref{pf-gks} is given by $\Delta=j+1$. The second is that the $\beta$-function 
vanishes for $\Delta=2$ which signals about massless modes. So putting these facts 
together, we recover the relation \eqref{kpz}.

(2) There is an equation in 2d quantum gravity that relates the gravitational scaling 
dimensions with the bare scaling dimensions \cite{gravrev}. It is the so-called 
Knizhnik-Polyakov-Zamolodchikov (KPZ) equation. Explicitly,
\begin{equation}\label{kpz-1}
\frac{\Delta^{\text{KPZ}}(\Delta^{\text{KPZ}}-1)}{\gamma_{\text s}-1}-
\Delta^{\text{KPZ}}=-\Delta^0
\quad.
\end{equation}
Here $\gamma_{\text s}$ means the string exponent (string susceptibility). 
In  the Liouville theory it is defined via the scaling of the partition function, i.e., 
$Z[A]\sim A^{\gamma_{\text s}-3}$, where $A$ is the invariant area. 

After this is noted, it immediately comes to mind to interpret the relation
\eqref{kpz} as an analogy of the KPZ equation. First, let us define the string exponent as 
$\hat Z[{\mathbf r}]\sim{\mathbf r}^{3-\gamma_{\text s}}$. A motivation for this 
definition is a simple analogy between ${\mathbf r}^{-1}$ and $A$ as they appear in 
the partition functions. It also assumes that  
$\int d^2z\,\beta\bbe\,\ep^{-\frac{2}{a}\varphi}$ can be interpreted as the 
cosmological operator in the problem of interest. Accepting the above definition, a result 
we can draw is $\gamma_{\text s}=2$. By substituting this value into 
Eq.\eqref{kpz-1}, we easily find the left hand side of the relation \eqref{kpz}. However, 
the missing point of the derivation sketched above is a possible interpretation of 
the bare dimensions as the masses of the effective scalars in $\ads$. 

Of course, these conclusions are heuristic and further work is needed to make them 
more rigorous.  

%_______________________      S E C T I O N - 4  _____________________

\section{ Conclusions and Remarks} 
\renewcommand{\theequation}{4.\arabic{equation}}
\setcounter{equation}{0}
First let us say a few words about the results.

In this work we have reproduced the basic features of the AdS/CFT correspondence for 
$\ads$ via the world-sheet methods and scaling arguments like in 2d gravity. 
In doing so, we 
proposed a stringy way for getting to a vicinity of critical points. In the case of interest we 
started from the 
world-sheet description of the CFT in space-time that is the boundary of $\ads$. 
Next we perturbed the world-sheet action by the marginal operator that gave us 
finite dimensional symmetry in space-time and provided the nonlinear term in the 
world-sheet action which correspond to $G_{\mu\nu}\pd X^\mu\pd X^\nu$ with 
the metric of $\ads$. We 
interpreted the result as string theory on $\ads$. 
The scale was introduced via the two-dimensional coupling constant
or, equivalently, the radial anti-de-Sitter coordinate. Then we studied the scaling 
limit via the 2d gravity scaling argument. We found the basic features of the AdS/CFT 
correspondence within our formalism. Although 
our procedure looks many ways attractive, we have to stress some its speculative 
character. It is clear that further work is needed to make this more rigorous. 

Let us conclude by mentioning a few problems that are seemed the most important to us.

(i)  In fact, all what we found in sections 2 and 3  corresponds to the effective action 
which is quadratic in the scalar field. It is known that such an action allows one to recover 
the conformal dimensions of the corresponding operators via their two-point functions. 
 So the open problem is to understand how to compute higher order terms in the 
effective action within our construction. 

(ii)  Relations between integrable Field Theories and Conformal Field Theories have been 
much studied of late. A possible approach to this problem by Zamolodchikov 
is to perturb CFT by some operators that lead to an integrable FT \cite{z}. 
The construction we proposed allows us to get to a vicinity of critical points in a stringy 
way. So it is rather natural to ask whether it may be used to study integrable Field 
Theories. 

(iii) Finally, the problem that obviously deserves more attention is supersymmetry 
breaking. For instance, $N=4\rightarrow N=2$ or $N=4\rightarrow N=0$.  
In the framework of our construction it means that we need to find such 
$\hat S_{\text{int}}$ that commutes only with the generators of the rest symmetry. 

%__________________       Acknowledgements   ________________________
\vspace{.25cm} {\bf Acknowledgments}

\vspace{.25cm} 
We benefited from discussions with H. Dorn, A. Fring, and 
I. Klebanov. It is also a pleasure to thank H. Dorn for reading the manuscript and 
useful comments.  We would like to acknowledge the hospitality of the Institut f\" ur Physik, 
Humboldt-Universit\"at zu Berlin, where this work was done. 
This research was supported in part by the Alexander von Humboldt Foundation
and by the European Community under Grant No. INTAS-OPEN-97-1312.

%__________________                      R E F S                    ______________________

%__________________________________________________________


\small
\begin{thebibliography}{99}
\bibitem{polbook} 
A. M. Polyakov, Gauge Fields and Strings, Hardwood Academic Publishers, 1987.
\bibitem{adsrev}
O. Aharony, S. Gubser, J. Maldacena, H. Ooguri, and Y. Oz, ``Large N Field 
Theories, String Theory and Gravity'', Report No. CERN-TH-99-122, 
hep-th/9905111.
\bibitem{mario}
P. M. Petropoulos, ``String Theory on $\ads$: Some Open Questions'', Report No. 
CPTH-PC-732.0899, hep-th/9908189.
\bibitem{malda} 
J. Maldacena, \ATMP{2}{1998} 231.
\bibitem{gkpw}
S. S. Gubser, I. R. Klebanov, and A. M. Polyakov, \PL{428}{1998} 105.  \\
E. Witten, \ATMP{2}{1998} 253.
\bibitem{bf}
P. Breitenlohner, and D. Z. Freedmann, Ann. Phys. 144 (1982) 249.
\bibitem{bkl} 
V. Balasubramanian, P. Kraus, and A. Lawrence, \PR{59}{1999} 046003.
\bibitem{kw}
I. R. Klebanov, and E. Witten, ``AdS/CFT Correspondence and Symmetry 
Breaking,'' Report No. PUPT-1863, hep-th/9905104.
\bibitem{k}
I. R. Klebanov, ``Absorption by Threebranes and the AdS/CFT Correspondence,''
presented at Strings '99, hep-th/9908165.
\bibitem{k2}
I. R. Klebanov, \PR{51}{1995} 1836; J. L. Barbon, K. Demeterfi, I. R. Klebanov, and 
C. Schmidhuber, \NP{440}{1995} 189.
\bibitem{ddk}
F. David, \MPL{3}{1988} 819; J. Distler, and H. Kawai, \NP{321}{1989} 509.
\bibitem{gravrev}
A. M. Polyakov, in Les Houches, 1988; I. R. Klebanov, Lectures at the Trieste 
Spring School, 1991; D. Kutasov, Lectures at the Trieste Spring School, 1991; 
F. David, in Les Houches, 1992.
\bibitem{gks} 
A. Giveon, D.  Kutasov, and N.  Seiberg, \ATMP{2}{1998} 733.  
\bibitem{andre1}
O. Andreev, \NP{552}{1999} 169.
\bibitem{h} 
G. 't Hooft, ``Dimensional Reduction In Quantum Gravity'',
  in Salamfest 1993, p.284.
\bibitem{suss} 
L. Susskind, J.Math.Phys. 36 (1995) 6377.
\bibitem{bpz} 
A. B. Belavin, A. M. Polyakov, and A. B. Zamolodchikov, \NP{241}{1984} 333.
\bibitem{d} 
Vl. S. Dotsenko, \NP{338}{1990} 747; B 358 (1991) 541.
\bibitem{andre2}
O. Andreev, ``Unitary Representations of Some Infinite Dimensional Lie Algebras 
Motivated by String Theory on $\ads$'', Report No. LANDAU-99/HEP-A2, 
hep-th/9905002, Nucl. Phys. B. to appear.
\bibitem{fms}
D. Friedan, E. Martinec, and S. Shenker, \NP{271}{1986} 93.
\bibitem{kt}
R. Kallosh, and A. Tseytlin, \HP{9810}{1998} 016.
\bibitem{pes}
I. Pesando, ``On the Quantization of the GS Type IIB Superstring Action on 
$\ads\times{\text S}^3$ with NSNS Flux'', Report No. DFTT-15-99, hep-th/9903089.
\bibitem{ger}
J.-L. Gervais, in Les Houches, 1995.
\bibitem{ds}
E. Diaconescu, and N. Seiberg, \HP{9707}{1997} 001.
\bibitem{sw} 
N. Seiberg, and E. Witten, \HP{9904}{1999} 017.
\bibitem{z}
A. B. Zamolodchikov, Adv. Stud. in Pure Math. 19 (1989) 641.
\end{thebibliography}
\end{document}